\documentclass[a4paper,11pt]{article}
\pdfoutput=1 

\usepackage{amsmath,amssymb}
\usepackage{array}
\usepackage{tabularx}
\usepackage{float}
\usepackage{jheppub} 

\usepackage[T1]{fontenc} 
\usepackage{tikz-cd}
\usepackage{caption}
\usepackage{booktabs} 

\title{\boldmath  Exact Stress Tensor Flows in Causal Electrodynamics: Auxiliary Field Formalism}

\author[*]{Zahra Pezhman,}
\author[\dag]{H. Babaei-Aghbolagh,}
\author[*]{Komeil Babaei Velni}

\affiliation[\dag]{Institute of Fundamental Physics and Quantum Technology, \& School of Physical Science and Technology, Ningbo University, Ningbo, Zhejiang 315211, China}
\affiliation[*]{Department of Physics, University of Guilan, P.O. Box 41335-1914, Rasht, Iran}
\emailAdd{ zpezhmanh@phd.guilan.ac.ir}
\emailAdd{hosseinbabaei@nbu.edu.cn}
\emailAdd{babaeivelni@guilan.ac.ir}

\abstract{Causal self-dual nonlinear electrodynamics can be described in terms of a single auxiliary field, with the theory specified by its potential \(W(\varphi)\). We use this formulation to study exact, nonperturbative \(T\bar T\)-like deformations. The irrelevant deformation rescales the auxiliary potential, whereas the marginal ModMax deformation shifts its argument. The resulting flow equations are obtained directly from the auxiliary-field action.
On any domain where $r(\psi)$ is regular and invertible, the irrelevant flow admits an autonomous representation, $\partial_\lambda \mathcal{L} = X F(r)$, with $r = -T^\mu{}_\mu/\sqrt{4X-Y}$ and $X, Y$ denoting the quadratic stress-tensor invariants. We derive the flow function \(F(r)\) and show how the auxiliary potential can be reconstructed from it. On an increasing, invertible branch, causality requires \(D(r)>1/(1-r)\). A globally single-valued flow function further requires the auxiliary-field ratio \(r(\psi)\) to remain one-to-one throughout the physical branch. We also study the relation between the $\varphi$-parity of the auxiliary potential, the evenness of \(F(r)\), and the analyticity of the Lagrangian, distinguishing the Maxwell and ModMax vacua. The construction extends to the Hamiltonian formulation, whose coupling flow has the opposite sign. We further express Legendre self-duality, the characteristic cones, and the energy conditions in terms of the flow function, and identify \(r=(\varepsilon-p_\perp)/(\varepsilon+p_\perp)\). The framework is illustrated by exact results for generalized Born--Infeld, logarithmic, \(q\)-deformed, and no-\(\tau\)-maximum electrodynamics.

}

\begin{document}
\maketitle
\flushbottom

\section{Introduction and main results}
\label{sec:intro}

The \(T\bar T\) deformation of two-dimensional quantum field theories \cite{Smirnov:2016lqw,Cavaglia:2016oda,Zamolodchikov:2004ce} (see \cite{Jiang:2019epa,He:2025ppz} for reviews) is a well-known example of a deformation generated by a composite operator constructed from the energy-momentum tensor. At the classical level, the flow equation \(\partial_\lambda\mathcal L=\det T_{\mu\nu}\) generates the Nambu--Goto action from a free scalar theory and provides a way to construct solvable irrelevant deformations. This approach has since been extended in two directions. First, the deformation operator can be generalized beyond the determinant of the stress tensor to functions of its Lorentz invariants. Such \(T\bar T\)-like deformations include the marginal root-\(T\bar T\) operator, which preserves classical integrability in sigma models \cite{Ferko:2022cix,Borsato:2022tmu,Babaei-Aghbolagh:2024hti}. Second, stress-tensor flows are not restricted to two dimensions and can be formulated in higher dimensions using metric or stress-tensor methods \cite{Taylor:2018xcy,Conti:2022egv,Blair:2024aqz,Brizio:2026ynw}. In four-dimensional nonlinear electrodynamics (NED), for example, Born--Infeld \cite{Born:1934gh} and ModMax theories can be obtained from Maxwell theory through such flows \cite{Babaei-Aghbolagh:2020kjg,Babaei-Aghbolagh:2022uij,Ferko:2022iru,Ferko:2023pr,Babaei-Aghbolagh:2024uqp}.


Four-dimensional nonlinear electrodynamics (NED) (see \cite{Sorokin:2021tge} for a review) includes an important class of theories invariant under electric--magnetic duality rotations \cite{Gaillard:1981rj,Gibbons:1995cv,Bialynicki-Birula:1984daz,Ivanov:2003uj}. For a Lagrangian \(\mathcal L(S,P)\), where \(S\) and \(P\) are the two electromagnetic invariants, duality invariance is equivalent to a nonlinear partial differential equation satisfied by both Born--Infeld and the conformal ModMax theory \cite{Bandos:2020jsw}. This property is preserved under stress-tensor flows \cite{Ferko:2023pr}. The relation between these flows and electric--magnetic duality, particularly for root-\(T\bar T\) deformations, has been studied in \cite{Babaei-Aghbolagh:2024uqp,Babaei-Aghbolagh:2025cni}, while their effects on birefringence were examined in \cite{Ferko:2023ruw}.

Russo and Townsend have also studied the conditions under which self-dual NED theories are causal, with electromagnetic disturbances propagating within or on the light cone, and the energy conditions they satisfy \cite{Russo:2024csd,Russo:2024xnh}. Such theories can be specified by a function of a single variable. In the Courant--Hilbert (CH) formulation, the Lagrangian is determined by a function \(\ell(\tau)\) subject to \(\dot\ell\geq1\) and \(\ddot\ell\geq0\) \cite{Russo:2024kto,Russo:2024dual,Courant:Hilbert}. Alternatively, Russo and Townsend introduced a formulation involving a single auxiliary field \(\varphi\) \cite{Russo:2025fuc} (see also \cite{Kuzenko:2025gvn,Kuzenko:2026hbw}), in which the Lagrangian takes the form
\[
\mathcal L(S,P;\varphi)
=\cosh\varphi\,S
+\sinh\varphi\,\sqrt{S^2+P^2}
-W(\varphi).
\]
Here \(W(\varphi)\) is the auxiliary potential and the equation of motion of \(\varphi\) is algebraic, \(\partial\mathcal L/\partial\varphi=0\). We use the conventions
$S=-\frac14F_{\mu\nu}F^{\mu\nu}, P=-\frac14F_{\mu\nu}\widetilde F^{\mu\nu}$.

Duality invariance is manifest in this formulation. Furthermore, strong-field causality is implied by weak-field causality for self-dual theories, and the causality and energy conditions can be expressed as inequalities involving \(W\) and its derivatives \cite{Russo:2025fuc,Russo:2024dual}.

Related auxiliary-field formulations have been used to construct integrable sigma models \cite{Ferko:2024ali,Bielli:2024ach,Fukushima:2024nxm,Fukushima:2025tlj,Fukushima:2026gan,Fukushima:2026znr,Bielli:2025uiv,Bielli:2026ggs,Baglioni:2026mii,Ferko:2026eci}. Their connections with four-dimensional duality-invariant electrodynamics and root-\(T\bar T\) flows have also been explored \cite{Baglioni:2025tsc,Babaei-Aghbolagh:2025uoz,Babaei-Aghbolagh:2025hlm}. This provides further motivation to study stress-tensor flows directly in terms of the auxiliary potential.

Both the Courant--Hilbert and auxiliary-field formulations describe self-dual nonlinear electrodynamics in terms of a function of one variable. Stress-tensor flows, however, are formulated in terms of quantities that depend on the electromagnetic fields. For theories with both an irrelevant deformation and a marginal ModMax deformation, it is therefore useful to relate the evolution of the defining function to the corresponding flow equations expressed through the stress-tensor invariants, \(X=T_{\mu\nu}T^{\mu\nu}\) and \(Y=(T^\mu{}_\mu)^2\), while keeping track of causality and the energy conditions.

Stress-tensor flows in the Courant--Hilbert formulation have been studied in \cite{Babaei-Aghbolagh:2025cni,Babaei-Aghbolagh:2024uqp,Babaei-Aghbolagh:2025hlm}. The classification of causal NED theories according to \(\varphi\)-parity and irrelevant deformations was initiated in \cite{Babaei-Aghbolagh:2026vkm}. We derived an exact nonperturbative \(T\bar T\)-like flow for one-scale self-dual NED theories admitting a Courant--Hilbert representation  \cite{Pezhman:2026sby,Pezhman:2026dsr}. Here we extend the analysis to the auxiliary-field formulation and investigate the reconstruction of the auxiliary potential, the Hamiltonian description, characteristic cones, and energy conditions.


We consider the two-parameter family of auxiliary potentials
\[
W(\varphi;\lambda,\gamma)=\frac{1}{\lambda}w(\varphi-\gamma),
\]
where \(\lambda\) is the irrelevant deformation coupling and \(\gamma\geq0\) is the marginal ModMax parameter. In this formulation, the two deformations act on the auxiliary potential through a rescaling and a shift of its argument. We relate the auxiliary potential to the Courant--Hilbert function and express the corresponding deformation equations in terms of the stress tensor. In particular, the irrelevant flow, on any domain
where $r(\psi)$ is regular and invertible, takes the form
\[
\partial_\lambda\mathcal L=X F(r),
\qquad
r=-\frac{T^\mu{}_\mu}{\sqrt{4X-Y}}=\frac{w}{w'},
\]
while the marginal flow is given by \(\partial_\gamma\mathcal L=\frac14\sqrt{4X-Y}\). We derive a parametric expression for \(F(r)\) and show how the auxiliary potential can be reconstructed from a prescribed flow function.

The auxiliary-field description also allows us to examine the conditions under which the flow is causal and globally single-valued. We express the causality constraints in terms of the potential and its derivatives, and obtain corresponding conditions on the flow function. An important distinction arises between the uniqueness of the auxiliary-field solution and the single-valuedness of \(F(r)\): causality  does not guarantee the latter, which requires another condition on the map between the auxiliary variable and \(r\). We then study the relation between reflection symmetry about the vacuum, \(\varphi\)-parity, and analyticity of the Lagrangian in the electromagnetic invariants $S$ and $P^2$. The role of the ModMax parameter is particularly relevant here, since these properties need not coincide when \(\gamma\neq0\). We also identify how odd terms in the expansion of \(F(r)\) affect analyticity in terms of $S$ and $P^2$ near the vacuum.

We give Lagrangian descriptions using angular, complex, and real auxiliary variables for the Maxwell-vacuum subclass with
$
\gamma=0,
$ $
w(-\psi)=w(\psi)
$, and extend the analysis to the Hamiltonian formulation, where the deformation equations acquire the opposite sign.  The flow function also provides a convenient way to study several physical properties of the resulting theories. In particular, we examine Born's Legendre self-duality, express the characteristic-cone coefficients in terms of \(F(r)\), and relate the flow variable to the energy density and transverse pressure through \(r=(\varepsilon-p_\perp)/(\varepsilon+p_\perp)\), in a frame where the electric and magnetic fields are parallel. We also show through $F(r)$ that a constant ratio of the two birefringence coefficients selects generalized Born--Infeld theory \cite{Russo:2025fuc}. Finally, we apply the construction to generalized Born--Infeld, logarithmic, \(q\)-deformed, and no-\(\tau\)-maximum electrodynamics, obtaining exact flow functions for these models.

The paper is organized as follows. In Sec.~\ref{sec:setup}, we introduce the auxiliary-field formulation and its relation to the Courant--Hilbert description. Section~\ref{sec:kernel} derives the exact \(T\bar T\)-like flow and the reconstruction formulas for the auxiliary potential. In Sec.~\ref{sec:global}, we examine the global properties of the flow function, focusing on its single-valuedness. Section~\ref{sec:parity} discusses \(\varphi\)-parity and analyticity. The linear auxiliary-field and Hamiltonian formulations are developed in Sec.~\ref{sec:hamiltonian}, while Sec.~\ref{sec:physics} examines duality, characteristic cones, and energy conditions. Exact examples are presented in Sec.~\ref{sec:models}. We conclude with a discussion in Sec.~\ref{sec:discussion}.


\section{Auxiliary fields and the Courant--Hilbert representation}\label{sec:setup}

We relate the auxiliary-field formulation of nonlinear electrodynamics
to the Courant--Hilbert (CH) description, in which the theory is encoded
by a function $\ell(\tau)$. The CH causality conditions are
\begin{equation}
 \dot\ell(\tau)\geq1,\qquad \ddot\ell(\tau)\geq0,
 \label{eq:CH-causality}
\end{equation}
where dots denote derivatives with respect to $\tau$.

\subsection{The Legendre map and causality conditions}

We define two exponential variables,
\begin{equation*}
 y=e^\varphi=e^{\gamma+\psi},\qquad
 \hat y=e^\psi.\qquad 
 \label{eq:y-yhat}
\end{equation*}
The corresponding potentials are defined by
\begin{equation*}
 \Omega(y)=\frac{1}{\lambda}w(\log y-\gamma),
 \qquad
 \widehat{\Omega}(\hat y)
 =\frac{1}{\lambda}w(\log\hat y).
 \label{eq:Omega-defs}
\end{equation*}
The two forms encode the same function
$w$, with the shift by $\gamma$ accounted for by $y=e^\gamma\hat y$.

The Legendre relations are
\begin{equation*}
 \ell(\tau)=\tau y-\Omega(y),\qquad
 \tau=\frac{d\Omega(y)}{dy},\qquad
 \dot\ell=y.
 \label{eq:Legendre}
\end{equation*}
Substituting $y=e^{\gamma+\psi}$, we obtain
\begin{equation}
 \tau(\psi)=\frac{e^{-(\gamma+\psi)}}{\lambda}w'(\psi),\qquad
 \ell\bigl(\tau(\psi)\bigr)=\frac{w'(\psi)-w(\psi)}{\lambda},
 \label{eq:tau-ell}
\end{equation}
where primes on $w$ denote derivatives with respect to $\psi$. It follows that
\begin{equation*}
 \tau\dot\ell=\frac{w'(\psi)}{\lambda},\qquad
 \ell-\tau\dot\ell=-\frac{w(\psi)}{\lambda}.
 \label{eq:CH-combinations}
\end{equation*}
Equivalently, the two representations satisfy
$\lambda\Omega(y)=\lambda\widehat{\Omega}(\hat y)=w(\psi)$,
with
$y\Omega'(y)=\hat y\widehat{\Omega}'(\hat y)
=w'(\psi)/\lambda$.


Differentiating $\tau(\psi)$ and using
$\dot\ell=e^{\gamma+\psi}$, we find
\begin{equation*}
 \frac{d\tau}{d\psi}
 =\frac{e^{-(\gamma+\psi)}}{\lambda}
 \bigl[w''(\psi)-w'(\psi)\bigr],
 \qquad
 \ddot\ell=
 \frac{\lambda e^{2(\gamma+\psi)}}
 {w''(\psi)-w'(\psi)}.
 \label{eq:ell-second}
\end{equation*}

The condition $\dot\ell\geq1$ follows from
$\dot\ell=e^{\gamma+\psi}$ on the domain $\gamma,\psi\geq0$.
 The remaining causality
condition $\ddot\ell>0$ requires
\begin{equation}
 \begin{cases}
 w''(\psi)>w'(\psi),&\lambda>0,\\
 w''(\psi)<w'(\psi),&\lambda<0.
 \end{cases}
 \label{eq:w-causality}
\end{equation}
These strict inequalities ensure  invertibility of the Legendre
map on any connected interval where they hold. The case $w''=w'$ where this map becomes degenerate requires separate treatment.

\subsection{Stress tensor flows }

To express the deformation flows in terms of the stress-tensor invariants, we first recall the energy-momentum tensor in the CH formulation. For $S^2+P^2>0$, the stress tensor takes the CH form
\begin{equation}
 T_{\mu\nu}
 =\frac{\tau\dot\ell}{\sqrt{S^2+P^2}}T^{\mathrm{Max}}_{\mu\nu}
 +(\ell-\tau\dot\ell)g_{\mu\nu}.
 \label{eq:stress-tensor}
\end{equation}
The Maxwell stress tensor is traceless and satisfies $
 T^{\mathrm{Max}}_{\mu\nu}T_{\mathrm{Max}}^{\mu\nu}
 =4(S^2+P^2).$
Using the relations derived above, we obtain
\begin{equation}
 X=\frac{4\bigl(w'(\psi)^2+w(\psi)^2\bigr)}{\lambda^2},\qquad
 Y=\frac{16w(\psi)^2}{\lambda^2}.
 \label{eq:stress-invariants}
\end{equation}
Consequently,
$4X-Y=16w'(\psi)^2/\lambda^2$.

Since the auxiliary-field equation of motion is satisfied, implicit derivatives of the auxiliary field do not contribute to the coupling derivatives of the on-shell Lagrangian.
We therefore obtain
\begin{equation}
 \partial_\lambda\mathcal L=\frac{w(\psi)}{\lambda^2},\qquad
 \partial_\gamma\mathcal L=\frac{w'(\psi)}{\lambda}.
 \label{eq:coupling-flows-parametric}
\end{equation}
On the branch where $w'(\psi)/\lambda>0$,
the marginal root-$T\bar T$ flow can be written as
\begin{equation}
 \partial_\gamma\mathcal L
 =\frac12\sqrt{X-\frac14Y}.
 \label{eq:gamma-flow}
\end{equation}

The coupling $\lambda$ rescales the auxiliary potential,
while $\gamma$ shifts its argument. On a branch
where the resulting Lagrangian and Hamiltonian are
sufficiently smooth functions of both couplings, the
corresponding flows commute:
\begin{equation*}
 [\partial_\lambda,\partial_\gamma]\mathcal L=0,
 \qquad
 [\partial_\lambda,\partial_\gamma]\mathcal H=0.
\end{equation*}

\section{Exact $T\bar T$-like flow and reconstruction of the potential}
\label{sec:kernel}
The trace relation alone does not provide an autonomous equation for the irrelevant flow, as its current expression still involves the coupling $\lambda$ explicitly. Our aim is to express $\partial_\lambda\mathcal L$ in terms of the stress-tensor invariants $X$ and $Y$. Using the relations  \eqref{eq:stress-invariants} and \eqref{eq:gamma-flow}, the trace and $\sqrt{4X-Y}$ are proportional to $w(\psi)$ and $w'(\psi)$, respectively. Their ratio is consequently independent of the overall normalization by $\lambda$ and of the shift by $\gamma$. We therefore introduce
\begin{equation}
    r=-\frac{T^\mu{}_\mu}{\sqrt{4X-Y}}
      =\frac{w(\psi)}{w'(\psi)},
    \label{eq:r}
\end{equation}
where the second equality holds on the branch
$\sqrt{4X-Y}=4w'(\psi)/\lambda>0$.

Substituting the expression for $X$ from
\eqref{eq:stress-invariants} into the
$\lambda$-flow equation in
\eqref{eq:coupling-flows-parametric}, we obtain
the exact irrelevant $T\bar T$-like flow in the
autonomous form
\begin{equation}
    \partial_\lambda\mathcal L=X\,F(r),
    \label{eq:flow}
\end{equation}
where the flow function is given parametrically by
\begin{equation}
    F\bigl(r(\psi)\bigr)
    =\frac{w(\psi)}
       {4\bigl[w'(\psi)^2+w(\psi)^2\bigr]}.
    \label{eq:kernel}
\end{equation}
Using the definition of $r$ in \eqref{eq:r},
the flow function can also be written as
\begin{equation}
    F\bigl(r(\psi)\bigr)
    =\frac{r(\psi)}
       {4w'(\psi)\bigl[1+r(\psi)^2\bigr]}.
    \label{eq:kernel-alt}
\end{equation}
The flow function $F$ is theory-dependent and has no
explicit dependence on $\lambda$ or $\gamma$; the coupling dependence enters through the stress-tensor invariant $X$ and the argument $r$.

\subsection{Reconstructing the auxiliary potential}

The flow function also determines the auxiliary potential. Suppose that
$r(\psi)$ is locally invertible, so that $\psi=\psi(r)$. Combining
$w=rw'$ with \eqref{eq:kernel}, we obtain
\begin{equation}
    w\bigl(\psi(r)\bigr)
    =\frac{r^2}{4(1+r^2)F(r)},
    \qquad
    w'\bigl(\psi(r)\bigr)
    =\frac{r}{4(1+r^2)F(r)}.
    \label{eq:wrec}
\end{equation}
Taking the logarithmic derivative of the first relation
in \eqref{eq:wrec} yields
\begin{equation*}
    \frac{d}{dr}\log
|w|
    =\frac{2}{r}-\frac{2r}{1+r^2}
      -\frac{F'(r)}{F(r)}.
\end{equation*}
Since $dw/d\psi=w'=w/r$, it follows that
\begin{equation}
    D(r)\equiv\frac{d\psi}{dr}
    =\frac{2}{1+r^2}
      -r\frac{F'(r)}{F(r)}.
    \label{eq:D}
\end{equation}
Integrating and imposing the vacuum condition $\psi(0)=0$ gives
\begin{equation}
    \psi(r)=2\arctan r
      -\int_{0}^r t\,\frac{F'(t)}{F(t)}\,dt.
    \label{eq:psirec}
\end{equation}
Thus, on the vacuum-connected branch where $r(\psi)$ is invertible and $F$ is regular, the flow function together with the vacuum condition determines $\psi(r)$ and $w(r)$,
and hence reconstructs the potential parametrically.

The vacuum behavior fixes the normalization locally. Assuming that the potential starts at quadratic order \cite{Pezhman:2026sby},
it means $w(0)=w'(0)=0$, so
 we have
\begin{equation*}
    r(\psi)=\frac{\psi}{2}+\mathcal O(\psi^2),
    \qquad D(0)=2,
\end{equation*}
and
\begin{equation*}
    w''(0)=\frac{1}{8F(0)}.
\end{equation*}
The normalization of $\lambda$ is conventional. Under the positive rescaling
\begin{equation*}
    \widetilde\lambda=c\lambda,\qquad c>0,
\end{equation*}
the same physical family is represented by
\begin{equation*}
    \widetilde w=cw,\qquad
    \widetilde F=\frac{F}{c}.
\end{equation*}
Indeed, $\widetilde w/\widetilde\lambda=w/\lambda$, while $\widetilde r=r$,
while the flow equation retains
its form, with
$\partial_{\widetilde\lambda}\mathcal L
=X\widetilde F(r)$.

\subsection{Causality constraints on the flow function}

We now express the CH causality condition in terms of
the flow function $F$.
We have
\begin{equation*}
\frac{w'}{\lambda}>0,\qquad \frac{w}{\lambda}>0,\qquad r>0.
\end{equation*}
Furthermore, the causality condition
\eqref{eq:w-causality} implies
\begin{equation*}
    \frac{d}{d\psi}\bigl(\frac{w'-w}{\lambda}\bigr)=\frac{w''-w'}{\lambda}>0, \qquad r<1.
\end{equation*}
A stronger bound follows by noting that
\begin{equation*}
    \frac{d}{dt}\bigl(\frac{e^{-t}w'(t)}{\lambda}\bigr)
    =e^{-t}\bigl(\frac{w''(t)-w'(t)}{\lambda}\bigr)>0.
\end{equation*}
Thus, for $0< t<\psi$, we have $\frac{w'(t)}{\lambda}
<e^{t-\psi}\frac{w'(\psi)}{\lambda}$.
Integrating from $0$ to $\psi$ and using $w(0)=0$,
we obtain
\begin{equation*}
    \frac{w(\psi)}{\lambda}=\int_{0}^\psi \frac{w'(t)}{\lambda}\,dt
    <\frac{w'(\psi)}{\lambda}\int_{0}^\psi e^{t-\psi}\,dt.
\end{equation*}
Consequently,
\begin{equation}
    r(\psi)<1-e^{-\psi},
    \qquad
    \psi(r)>-\log(1-r).
    \label{eq:r-bound}
\end{equation}
This bound imposes a constraint on the stress-tensor
ratio $r$ along causal flows. Differentiating
$r=w/w'$ gives
\begin{equation*}
    \frac{dr}{d\psi}=1-r\frac{w''}{w'}.
\end{equation*}
On a branch where $r(\psi)$ is invertible,
we use $D=d\psi/dr$ to obtain
\begin{equation}
    \frac{w''-w'}{w'}
    =\frac{(1-r)D-1}{rD}.
    \label{eq:caus-D}
\end{equation}
On the increasing branch connected continuously to the vacuum,
$D>0$. For $0<r<1$, the sign of $\frac{w''-w'}{w'}$ is therefore the sign of $(1-r)D-1$. Hence
\begin{equation}
    D(r)>\frac{1}{1-r}.
\end{equation}
Substituting \eqref{eq:D}, we obtain the differential inequality
\begin{equation}
    \frac{2}{1+r^2}
    -r\frac{F'(r)}{F(r)}
    >\frac{1}{1-r}.
    \label{eq:causK}
\end{equation}
On the regular increasing branch with $0<r<1$,
this inequality is the necessary and sufficient
condition on $F(r)$ for CH causality.

\section{Global structure: single-valuedness and principal invariant}
\label{sec:global}

The reconstruction in Sec.~\ref{sec:kernel} assumes that $r(\psi)$ is locally invertible. Whether the closed form $F(r)$ is a single function on the whole physical branch is a global question, to which causality alone does not provide the answer. Causality ensures a unique solution
to the auxiliary-field equation for each admissible electromagnetic configuration, since it makes $\Omega''>0$; it does not, however, guarantee that
$r=w/w'$ is globally injective.

The local reconstruction in Sec.~\ref{sec:kernel} treats $r$ as a  coordinate on the auxiliary-field branch. For $F$ to be globally well-defined as a function of $r$, distinct values of $\psi$ must not map to the same value of $r$. This requirement is  independent of the local CH causality condition.

Differentiating $r=w/w'$ gives $dr/d\psi=1-rw''/w'$. The flow function $F(r)$ is globally single-valued on a physical interval provided that
\begin{equation*}
    w'^2-ww''>0,
\end{equation*}
throughout that interval, or equivalently,
$dr/d\psi>0$.
Under causality and this additional condition, we have
\begin{equation}\label{ci}
    1<\frac{w''}{w'}<\frac{1}{r}.
\end{equation}
Thus, $F(r)$ is globally single-valued on the image
of the physical interval.

Conversely, we have $\lambda F(r)>0$ for
$0<r<r_*$, with $r_*<1$, and $F(r)$ is analytic
near $r=0$ (in the sense that it admits a Taylor
expansion), with $F(0)\neq0$, and sufficiently
differentiable elsewhere. If $D(r)>1/(1-r)$ throughout this interval, the reconstruction formulas define a unique causal potential on the corresponding $\psi$-interval.

Global single-valuedness imposes an additional restriction. Since $w''/\lambda>w'/\lambda>0$, the function $w/\lambda$ is strictly increasing. Thus, if two distinct values of $\psi$ correspond to the same
positive value of $r$, the corresponding values of
\begin{equation*}
    F=\frac{r}{4w'(1+r^2)}
\end{equation*}
would differ. The flow function would therefore not
be globally single-valued on the physical branch.
More precisely, at a point where $dr/d\psi=0$,
\begin{equation*}
    \frac{dF}{d\psi}=-F\,\frac{w''}{w'}\neq0,
\end{equation*}
whereas a differentiable function $F(r)$ would satisfy
$dF/d\psi=F'(r)\,dr/d\psi=0$.
Hence, $F$ cannot be represented as a differentiable
function of $r$ through such a point. The obstruction
arises because the stress-tensor ratio $r$ ceases to
provide a local coordinate on the
auxiliary-field branch.

The ratio $r$ is useful for reconstructing the potential, but its definition involves a square root of the stress-tensor invariants. We therefore introduce the
dimensionless ratio $\rho=\sqrt{Y/X}$, defined directly in terms of $X$ and $Y$.

The two dimensionless ratios are related by
\begin{equation}\label{rhor}
\rho=\frac{2r}{\sqrt{1+r^2}},
\qquad
r=\frac{\rho}{\sqrt{4-\rho^2}}.
\end{equation}
To distinguish the two variables, we define $\mathcal{F}(\rho)=F(\rho/\sqrt{4-\rho^2})$. The exact autonomous flow then takes the form
\begin{equation}
    \partial_\lambda\mathcal{L}
    =X\,\mathcal{F}(\rho).
    \label{eq:flowrho}
\end{equation}

The causal bound $r<1$ translates into $0<\rho<\sqrt{2}$, away from degenerate zero-invariant configurations. This interval corresponds to the physical branch expressed in terms of the stress-tensor invariants.

If $\mathcal F$ is analytic near the vacuum (again in the sense of a Taylor expansion),
\begin{equation*}
\mathcal{F}(\rho)=\sum_{m=0}^{\infty}c_m\rho^m,
\end{equation*}
the flow becomes
\begin{equation}
\partial_\lambda\mathcal{L}
=\sum_{m=0}^{\infty}
c_m\,Y^{m/2}X^{1-m/2}.
\end{equation}
Odd powers of $\rho$ give rise to half-integer powers of $X$ and $Y$. Evenness of $\mathcal{F}(\rho)$ removes these half-integer powers, but does not by itself make the operator polynomial in $X$ and $Y$. For example, a $\rho^4$ term produces $Y^2/X$. This shows that analyticity in the electromagnetic invariants $S,P^2$ and analyticity as a function of independently varied stress-tensor invariants are different properties.

The evenness of the flow function $F(r)$ is related to that of the auxiliary potential $w(\psi)$ through
\begin{equation}
    F(-r)=F(r)
    \quad\Longleftrightarrow\quad
    w(-\psi)=w(\psi).
\end{equation}
Indeed, if $w$ is even, then $w'$ and $r=w/w'$ are odd, while the parametric expression for $F$ is even. Conversely, if $F(r)$ is even, then $D(r)$ is even, $\psi(r)$ is odd, and the reconstructed potential
\begin{equation*}
    w(r)=\frac{r^2}{4(1+r^2)F(r)}
\end{equation*}
is even. Here negative $r$ refers to the signed analytic continuation of $w/w'$, not to a negative value of $\rho$.


\section{$\varphi$-Parity and analyticity}
\label{sec:parity}


  The statement that $F$ is even should be distinguished
from other symmetry properties. In this section, we
consider three related but distinct notions: reflection
about the selected vacuum, $\varphi$-parity,
and analyticity of the Lagrangian in the electromagnetic
invariants $S$ and $P^2$. The distinction depends on
the marginal parameter $\gamma$, which shifts the
auxiliary-field vacuum away from $\varphi=0$.

Let $w^{\mathrm c}(\psi)=w(-\psi)$. Then vacuum-centred reflection acts on the data as
\begin{equation}
(w,\gamma)
\longmapsto
\bigl(w^{\mathrm c},\gamma\bigr), \qquad r^{\mathrm c}(\psi)=-r(-\psi).
\end{equation}
This reverses \(\psi\) while keeping the vacuum fixed.
The corresponding flow function satisfies
$F^{\mathrm c}(r)=F(-r)$. If $F(r)$ is analytic
near $r=0$, vacuum-centred reflection reverses
the signs of its odd Taylor coefficients.

$\varphi$-parity, associated with the sign
reversal of the square-root invariant $\sqrt{S^2+P^2}$,
instead acts on the \emph{unshifted} potential as
${\widehat{W}}(\varphi)=W(-\varphi)$.


An even $w(\psi)$ expresses symmetry about the selected vacuum $\varphi=\gamma$. This symmetry corresponds to invariance of the vacuum-connected theory under $\varphi$-parity only when $\gamma=0$. For $\gamma>0$, the $\varphi$-parity dual has a ModMax vacuum with parameter $-\gamma$, which violates causality. For $\gamma=0$, however, both members of a
$\varphi$-parity pair can be causal, although they
need not coincide. Thus, symmetry of the potential about its vacuum and invariance under  $\varphi$-parity are distinct notions that coincide only when $\gamma=0$.

The distinction between vacuum-centred reflection and $\varphi$-parity becomes especially clear in the
weak-field expansion. The leading term determines
whether nonanalytic dependence on the electromagnetic
invariants is already present before the irrelevant
deformation is switched on.

To examine this distinction at the level of the Lagrangian, we expand around the vacuum. The weak-field limit is
\begin{equation}
\mathcal{L}_{\mathrm{weak}}
=S\cosh\gamma+\sqrt{S^2+P^2}\sinh\gamma.
\end{equation}
Define the combination
\begin{equation*}
J_\gamma
=S\sinh\gamma+\sqrt{S^2+P^2}\cosh\gamma
=e^\gamma V+e^{-\gamma}U,
\end{equation*}
where
\[
U = \frac{\sqrt{S^{2}+P^{2}} - S}{2},
\qquad
V = \frac{\sqrt{S^{2}+P^{2}} + S}{2}.
\]
Solving the auxiliary equation near the vacuum gives
\begin{equation*}
\psi=\frac{\lambda J_\gamma}{w''(0)}+O(\lambda^2).
\end{equation*}
Expanding the Lagrangian to first order in \(\lambda\) and then substituting the result obtained above into the expansion, we obtain
\begin{equation}
\mathcal{L}
=\mathcal{L}_{\mathrm{weak}}
+\frac{\lambda}{2w''(0)}J_\gamma^2
+O(\lambda^2).
\end{equation}

For $\gamma>0$, nonanalytic dependence on the electromagnetic invariants is already
present in the leading ModMax term, because the
square-root invariant $\sqrt{S^2+P^2}$ appears linearly there;
evenness of the irrelevant flow function $F$
cannot remove it. For $\gamma=0$ and in the weak-field limit, the regular
analytic reconstruction gives
\begin{equation}
\begin{aligned}
{\mathcal{L}\ \text{analytic in }(S,P^2)}
&\;\Longleftrightarrow\;
\varphi\text{-parity invariance} 
&\;\Longleftrightarrow\;
 w\text{ even}
\;\Longleftrightarrow\;
F\text{ even}\;.
\end{aligned}
\label{eq:analyticity-parity-kernel}
\end{equation}

For $\gamma=0$, expanding the Lagrangian to
second order in $\lambda$ gives

\begin{equation}
\begin{aligned}
\mathcal{L}
={}&S
+\frac{\lambda}{2w''(0)}(S^2+P^2)
+\frac{\lambda^2}{2w''(0)^2}S(S^2+P^2)\\
&-\frac{w'''(0)\lambda^2}{6w''(0)^3}
(S^2+P^2)^{3/2}
+O(\lambda^3).
\end{aligned}
\end{equation}

The first odd coefficient in the expansion of $F(r)$
therefore characterizes the leading nonanalytic interaction:
\begin{equation}
F'(0)=-\frac{w'''(0)}{6w''(0)^2}.
\end{equation}
The condition \(F'(0)=0\) removes this particular obstruction, whereas full analyticity of \(\mathcal{L}\) in \(S\) and \(P^2\) near the Maxwell vacuum requires the vanishing of all odd-order Taylor coefficients of \(F\) at \(r=0\).

\begin{table}[H]\centering
\small
\setlength{\tabcolsep}{3pt}
\renewcommand{\arraystretch}{1.25}

\begin{tabular}{
  p{0.25\linewidth}
  p{0.32\linewidth}
  p{0.32\linewidth}
}
\hline
\textbf{Statement}
& \textbf{Maxwell vacuum, $\gamma=0$}
& \textbf{ModMax vacuum, $\gamma>0$} \\
\hline

Even regular flow function $F$
& Potential symmetric about the vacuum
& Potential symmetric about the vacuum \\[5pt]

 $\varphi$-parity invariance
& Equivalent to evenness of the potential 
& Changes the selected vacuum to $-\gamma$ \\[5pt]

Analyticity in $S,P^2$
& Equivalent to an even analytic F
& Obstructed by the leading ModMax term \\


\hline
\end{tabular}
\caption{$\varphi$-Parity, vacuum and analyticity statements for the reconstructed class.}\label{tab:parity}
\end{table}


\section{Linear auxiliary and Hamiltonian formulations}
\label{sec:hamiltonian}

The reconstruction in Secs.~\ref{sec:kernel}--\ref{sec:parity} defines the theory through the potential $w$ and the auxiliary field. We now show that, for the Maxwell-vacuum class with \(\gamma=0\) and \(w(-\psi)=w(\psi)\), which is analytic in \(S\) and \(P^2\), the same theory admits formulations in which the Lagrangian is linear in \(S\) and \(P\). The corresponding Hamiltonian also satisfies the flow equation of Sec.~\ref{sec:kernel}, but with the opposite sign.

For the Maxwell-vacuum case, evenness of $w$ allows the sign ambiguity arising from the elimination of the angular auxiliary field to be absorbed by reflecting $\psi$. In these formulations, $S,P$ enter linearly, while all nonlinear information is contained in the auxiliary potential.

Introduce an angular auxiliary field $\vartheta$ in addition to $\psi$:

\begin{equation}\label{Llinear}
\mathcal L
=\left(\cosh\psi+\sinh\psi\cos\vartheta\right)S
+\sinh\psi\sin\vartheta\,P
-\frac{w(\psi)}{\lambda},
\end{equation}
where
\begin{equation}\nonumber
    \cos\vartheta = \pm \frac{S}{\sqrt{S^2 + P^2}},
    \qquad
    \sin\vartheta = \pm \frac{P}{\sqrt{S^2 + P^2}}.
\end{equation}
This gives a definite Lagrangian linear in $S,P$.
Eliminating $\vartheta$ gives

\begin{equation}
\mathcal L
=S\cosh\psi\pm\sqrt{S^2+P^2}\sinh\psi
-\frac{w(\psi)}{\lambda}.
\end{equation}

The two signs are equivalent, since $w$ is even. To express its potential as an analytic function of a complex auxiliary variable, we introduce transformed potential
\begin{equation}
\begin{aligned}
z&=\sinh\psi\,e^{i\vartheta},
\qquad t=z\bar z,\\
w_z(t)&:=w\!\left(
\operatorname{arsinh}\sqrt{t}
\right).
\end{aligned}
\end{equation}

Since $w$ is analytic and even, $w_z$ is analytic near $t=0$. The Lagrangian can then be written as
\begin{equation}
\mathcal L
=\left[\sqrt{1+z\bar z}+\Re z\right]S
+(\Im z)P
-\frac{w_z(z\bar z)}{\lambda}.
\end{equation}

The flow function $F(r)$ determines the transformed
potential parametrically:
\begin{equation}
t(r)=\sinh^2\psi(r),
\qquad
w_z\bigl(t(r)\bigr)
=\frac{r^2}{4(1+r^2)F(r)}.
\end{equation}

Formulation \eqref{Llinear} uses two independent real
auxiliary fields $a>0$ and $b$, with
\begin{equation}
c(a,b)
=\frac12\left[a+\frac{1+b^2}{a}\right],
\qquad
\mathcal U(c)
=w\!\left(\operatorname{arcosh}c\right),
\end{equation}
\begin{equation*}
a
=\left(\cosh\psi+\sinh\psi\cos\vartheta\right),\qquad
b=\sinh\psi\sin\vartheta.
\end{equation*}

The Lagrangian then takes the form
\begin{equation}
\mathcal L
=aS+bP
-\frac{\mathcal U(c(a,b))}{\lambda}.\label{eq:lin}
\end{equation}

In this formulation, duality is particularly transparent, as will be  discussed in Sec.~\ref{sec:physics}. The corresponding Hamiltonian is
\begin{equation}\label{Hlinear}
\mathcal H
=\frac{\lvert\mathbf D-b\mathbf B\rvert^2}{2a}
+\frac a2\lvert\mathbf B\rvert^2
+\frac{\mathcal U(c(a,b))}{\lambda}.
\end{equation}

The preceding linear formulation \eqref{Hlinear} applies without a
sign ambiguity to the analytic, $\varphi$-parity-invariant class. 

For a general potential, which need not be even, we instead work directly with the Hamiltonian variables. Define
\begin{equation}
s=\frac12\left(|\mathbf D|^2+|\mathbf B|^2\right),
\qquad
p=|\mathbf D\times\mathbf B|,
\label{eq:ham-sp}
\end{equation}
and
\begin{equation}
u=\frac{s-\sqrt{s^2-p^2}}{2},
\qquad
v=\frac{s+\sqrt{s^2-p^2}}{2}.
\end{equation}

The corresponding auxiliary Hamiltonian is
\begin{equation}
\mathcal H_{\mathrm{aux}}
=e^{\psi+\gamma} u+e^{-(\psi+\gamma)}v+\frac{w(\psi)}{\lambda}.
\end{equation}
Its field equation is
\begin{equation*}
\frac{w'}{\lambda}
=e^{-(\psi+\gamma)}v-e^{\psi+\gamma} u.
\end{equation*}
Equivalently, the auxiliary-field equation can be written as
\begin{equation}
\Omega'(y)=-u+\frac{v}{y^2}.
\end{equation}
On the physical branch, where $\Omega''(y)>0$, this
equation has at most one solution. At such a solution,
\begin{equation*}
v\geq e^{2(\psi+\gamma)}u,
\end{equation*}
or, for $\psi+\gamma\geq0$, equivalently,
\begin{equation*}
s\geq p\cosh(\psi+\gamma).
\end{equation*}

The Hamiltonian Courant--Hilbert boundary function is reconstructed by
\begin{equation}
\sigma(\psi)
=\frac{e^{\gamma+\psi}w'(\psi)}{\lambda},
\qquad
h\bigl(\sigma(\psi)\bigr)
=\frac{w'(\psi)+w(\psi)}{\lambda}.
\end{equation}
The flow function $F(r)$ gives the boundary data
of the Lagrangian and Hamiltonian Courant--Hilbert
descriptions explicitly:
\begin{equation}
\tau(r)
=\frac{e^{-\gamma-\psi(r)}r}
       {4\lambda(1+r^2)F(r)}, \qquad \sigma(r)
=\frac{e^{\gamma+\psi(r)}r}
       {4\lambda(1+r^2)F(r)},
\end{equation}
and
\begin{equation}
\ell(r)
=\frac{r(1-r)}{4\lambda(1+r^2)F(r)},
\qquad
h(r)
=\frac{r(1+r)}{4\lambda(1+r^2)F(r)}.
\end{equation}
Their derivatives satisfy
\begin{equation}
h'=e^{-(\psi+\gamma)},
\qquad
h''=-\frac{\lambda e^{-2(\psi+\gamma)}}{w''+w'},
\end{equation}
and
\begin{equation}
h'+2\sigma h''
=e^{-(\psi+\gamma)}\frac{w''-w'}{w''+w'}.
\end{equation}
Hence the Hamiltonian causality conditions follow directly from the same reconstructed potential: for $\frac{w''}{\lambda}>\frac{w'}{\lambda}>0$ one has $\frac{w''+w'}{\lambda}>0$, so $h'+2\sigma h''$ is positive.

Finally, the coupling derivatives of the Hamiltonian at fixed $\mathbf D,\mathbf B$ are
\begin{equation}
\partial_\lambda\mathcal H
=-\frac{w}{\lambda^2}
=-X F(r), \qquad \partial_\gamma\mathcal H
=-\frac{w'}{\lambda}
=-\frac14\sqrt{4X-Y}.\label{eq:dlH}
\end{equation}
The Lagrangian and Hamiltonian coupling derivatives therefore have opposite signs, while involving the same stress-tensor invariants and the same flow function $F(r)$.


\section{Duality, characteristic cones and energy conditions}
\label{sec:physics}

We now draw the physical consequences of the reconstructed potential. We first establish the duality properties, then study the propagation of perturbations on a static background in terms of the two birefringence coefficients $A_\pm$, and finally relate the flow variable $r$ to the energy density and pressures.

The auxiliary formulation also makes it possible to separate properties that are sometimes conflated. Electromagnetic self-duality is a continuous duality property of the equations, Legendre self-duality is a property of the Hamiltonian under a discrete Legendre transformation, and  $\varphi$-parity is an additional reflection symmetry. The first two need not imply the third.

Every reconstructed theory remains electromagnetically self-dual because the Lagrangian satisfies
\begin{equation}
\mathcal{L}_U \mathcal{L}_V = -1,
\end{equation}

with $U,V=(\sqrt{S^2+P^2}\mp S)/2$. This property follows from the structure of $\mathcal L$ as a linear function of $S$ and $\sqrt{S^2+P^2}$, with a duality-invariant potential. Legendre self-duality follows from the discrete duality rotation
\begin{equation}
\mathcal{H}(\mathbf{D},\mathbf{B})
=
\mathcal{H}(\mathbf{B},-\mathbf{D}).
\end{equation}

It therefore holds for the reconstructed theory irrespective of whether $F$ is even. For the even-potential class, the result can also be seen directly in the linear auxiliary formulation under
\begin{equation}
\widetilde{a} = \frac{a}{a^2+b^2},
\qquad
\widetilde{b} = -\frac{b}{a^2+b^2},
\end{equation}

one has
\begin{equation*}
c(\widetilde{a},\widetilde{b}) = c(a,b).
\end{equation*}

The reconstructed interaction potential is therefore invariant under the auxiliary redefinition implementing the covariant Legendre transformation. Electromagnetic self-duality,  Legendre self-duality and $\varphi$-parity under $\varphi\mapsto-\varphi$ are distinct statements: the first two hold throughout the reconstructed class, while the third imposes the additional $\varphi$-parity condition of Sec.~\ref{sec:parity}.

Causality can be tested not only through convexity of the CH function, but also through the propagation of small disturbances on a background electromagnetic field. In a frame where the background electric and magnetic fields are parallel, the two polarizations have dispersion relations controlled by two coefficients, which can be expressed directly in terms of the reconstructed potential.

For plane-wave perturbations about a static background, the two dispersion relations are
\begin{equation}
\omega^2
=
k_\parallel^2+A_\pm k_\perp^2,
\end{equation}
where the birefringence coefficients are given in
terms of the auxiliary potential by
\begin{equation}
A_-=e^{-2(\psi+\gamma)},
\qquad
A_+=\frac{w''-w'}{w''+w'}.\label{eq:Apm}
\end{equation}

Using $w''/w'=(D-1)/(rD)$, the exact flow function
$F(r)$ determines the birefringence coefficients as
\begin{equation}
A_-(r)=e^{-2\gamma-2\psi(r)},
\qquad
A_+(r)=
\frac{(1-r)D(r)-1}{(1+r)D(r)-1}.
\end{equation}

The causality condition on the flow function implies
\begin{equation*}
0<A_+\leq 1,
\qquad
0<A_-\leq 1.
\end{equation*}

At the weak-field limit,

\begin{equation}
A_+(0)=1,
\qquad
A_-(0)=e^{-2\gamma}.
\end{equation}

Thus $F(r)$ determines $A_+$ through $D(r)$, while $A_-$ also requires the reconstructed function $\psi(r)$, with $\gamma$ fixing its weak-field value. The ratio is
\begin{equation*}
\frac{A_-}{A_+}
=
e^{-2\gamma-2\psi(r)}
\frac{(1+r)D(r)-1}{(1-r)D(r)-1}.
\end{equation*}

It is natural to ask when this ratio stays at its vacuum value along the whole flow, i.e.\ when the two cones are related by a constant factor. Requiring
\begin{equation*}
\frac{A_-}{A_+}=e^{-2\gamma},
\end{equation*}

for all $\psi$ is equivalent to
\begin{equation*}
\frac{w''}{w'}=\coth\psi.
\end{equation*}

The solution is
\begin{equation}
w(\psi)=w''(0)(\cosh\psi-1).\label{eq:gbiw}
\end{equation}

Consequently, the corresponding flow function is
\begin{equation}
\begin{aligned}
F(r)
&=\frac{1-r^2}{8w''(0)(1+r^2)},\\
\mathcal F(\rho)
&=\frac{1}{8w''(0)}
-\frac{\rho^2}{16w''(0)}.
\end{aligned}
\end{equation}
This identifies generalized Born--Infeld electrodynamics
(Sec.~\ref{sec:models}): a constant ratio $A_-/A_+$
singles out the generalized Born--Infeld family.

We now examine the energy density and principal pressures to clarify the physical meaning of the auxiliary ratio $r$ and establish the dominant and strong energy conditions without specifying a particular model. Causality implies
\begin{equation*}
\frac{w'}{\lambda}>\frac{w}{\lambda}>0
\qquad (\psi>0).
\end{equation*}
This inequality ensures that the null, weak, dominant, and strong energy conditions hold.
To make these conditions explicit, consider a non-null electromagnetic field and choose a frame in which the electric and magnetic fields are parallel, with $\vec E=(0,0,E)$ and $\vec B=(0,0,B)$. The energy density and principal pressures are
\begin{equation}
\varepsilon=\frac{w'+w}{\lambda},
\qquad
p_{\perp}=\frac{w'-w}{\lambda},
\qquad
p_{\parallel}=-\frac{w'+w}{\lambda}.
\end{equation}

Writing $A=w'/\lambda$, these expressions become
\begin{equation}
\varepsilon=A(1+r),
\qquad
p_{\perp}=A(1-r),
\qquad
p_{\parallel}=-\varepsilon.
\end{equation}

The ratio $r$ therefore has the invariant physical interpretation
\begin{equation}
r=\frac{\varepsilon-p_{\perp}}
        {\varepsilon+p_{\perp}}.
\end{equation}
Since $0<r<1$, we have
\begin{equation*}
\varepsilon>0,
\qquad
0< p_{\perp}<\varepsilon,
\qquad
p_{\parallel}<0.
\end{equation*}
Moreover,
\begin{equation*}
\varepsilon+p_{\perp}=2A>0,
\qquad
\varepsilon+p_{\parallel}=0,
\end{equation*}
and
\begin{equation*}
\varepsilon+2p_{\perp}+p_{\parallel}
=2A(1-r)>0.
\end{equation*}

Thus, the null, weak, dominant, and strong energy conditions are satisfied. The flow function $F(r)$ then reconstructs the energy density and transverse pressure through
\begin{equation}
\begin{aligned}
A(r)&=\frac{r}{4\lambda(1+r^2)F(r)},\\
\varepsilon(r)
&=\frac{r(1+r)}{4\lambda(1+r^2)F(r)},\\
p_{\perp}(r)
&=\frac{r(1-r)}{4\lambda(1+r^2)F(r)}.
\end{aligned}
\end{equation}

Finally, the traceless part of the mixed stress tensor, $\hat T^\mu{}_\nu=T^\mu{}_\nu-\tfrac{1}{4}T^\alpha{}_\alpha\delta^\mu{}_\nu$, is proportional to the Maxwell stress tensor and therefore satisfies
\begin{equation}
\hat T^\mu{}_\rho\,
\hat T^\rho{}_\nu
=\frac{4X-Y}{16}\,\delta^\mu{}_\nu.
\end{equation}

Consequently, all algebraic scalar invariants of $T^\mu{}_\nu$ can be expressed in terms of its trace and $X$. This provides an algebraic explanation for the closure of the flow on these two invariants.


\section{Exact models}
\label{sec:models}
In this section, we apply the reconstruction to representative examples from standard families of causal self-dual electrodynamics. For each theory, we start from the CH boundary function $\ell(\tau)$ and obtain the auxiliary potential through $w=\lambda(\tau\dot\ell-\ell)$. After checking causality through $\frac{w''-w'}{\lambda}$, we determine $r(\psi)=w/w'$ and invert this relation to obtain $\psi(r)$. Substitution into the reconstruction formula then yields the flow function $F(r)=r/[4w'(1+r^2)]$.

The inversion is the potentially nontrivial step, which may require solving an algebraic equation or using a more general inverse function. It is also at this stage that the global single-valuedness criterion of Sec.~\ref{sec:global} becomes relevant. The invariant flow function $\mathcal F(\rho)$ then follows from the universal change of variable in \eqref{rhor}.

 \subsection{Generalized Born--Infeld}
For generalized Born--Infeld electrodynamics, we start directly from the CH function
\begin{equation}
\ell_{\mathrm{GBI}}(\tau)
=\frac{1-\sqrt{1-2e^\gamma\lambda\tau}}{\lambda}.
\end{equation}

The corresponding auxiliary potential is $w_{\mathrm{GBI}}(\psi)=\cosh\psi-1$. Thus,
\begin{equation*}
r(\psi)=\frac{\cosh\psi-1}{\sinh\psi}
=\tanh\frac{\psi}{2},
\qquad
\psi(r)=2\operatorname{arctanh}r= \ln(\frac{1+r}{1-r}).
\end{equation*}

The corresponding exact flow function and its invariant representation are
\begin{equation}
\begin{aligned}
F_{\mathrm{GBI}}(r)
&=\frac{1-r^2}{8(1+r^2)},\\
\mathcal F_{\mathrm{GBI}}(\rho)
&=\frac18-\frac{\rho^2}{16}.
\end{aligned}
\end{equation}

The resulting exact flow takes the explicit form
\begin{equation}
\partial_\lambda\mathcal L_{\mathrm{GBI}}
=X\,\mathcal F_{\mathrm{GBI}}
\left(\sqrt{\frac{Y}{X}}\right)
=\frac{X}{8}-\frac{Y}{16}.
\end{equation}
As shown in Sec.~\ref{sec:physics}, this is the unique family for which the ratio $A_-/A_+$ remains constant along the flow.

\subsection{Logarithmic theory}
For the logarithmic theory,  the CH function is
\begin{equation}\nonumber
    \ell_{\log}(\tau)
    = -\frac{1}{\lambda}
      \log\left(1-e^\gamma\lambda\tau\right).
\end{equation}

The corresponding auxiliary potential is
\begin{equation}\nonumber
   w_{\log}(\psi)=e^\psi-1-\psi.
\end{equation}

Setting $s=e^{-\psi}$, the auxiliary potential
and its derivative become
\begin{equation*}
\begin{aligned}
w&=\frac{1}{s}-1+\log s,
\qquad
w'=\frac{1-s}{s},\\
r(s)&=1+\frac{s\log s}{1-s}.
\end{aligned}
\end{equation*}

Inverting this relation with the condition $1-\psi<e^{-\psi}$, gives
\begin{equation}
s(r)
=\frac{r-1}{W_{-1}((r-1)e^{r-1})},
\qquad 0<r<1.
\end{equation}
Here $W_{-1}$ denotes the lower real branch of the Lambert function, which gives the nontrivial physical solution $0<s<1$. The principal branch instead yields $s=1$, which is not a solution of the original relation for $r>0$. The vacuum value $s(0)=1$ is recovered by continuity.

The resulting flow function is
\begin{equation}
F_{\log}(r)
=\frac{r\,s(r)}
       {4[1-s(r)](1+r^2)}.
\end{equation}
Since $w''-w'=1$, to establish causality, we need to take $\lambda$ as positive throughout the physical auxiliary interval.
Unlike the generalized Born--Infeld model, the
logarithmic potential is not even, and the resulting
flow function contains odd powers of $r$.
In particular, the relation derived in
Sec.~\ref{sec:parity} gives
$F'_{\log}(0)=-1/6$.


\subsection{The $q$-deformed family}
The $q$-deformed family is specified by the
Courant--Hilbert function
\begin{equation}
\ell_q(\tau)
=
\frac{
1-\left(1-\frac{e^\gamma\lambda\tau}{q}\right)^q
}{\lambda},
\qquad q>0.
\end{equation}

For $q\neq 1$, differentiating the CH boundary function gives $\ell_q'(\tau)=e^\gamma (1-e^\gamma\lambda\tau/q)^{q-1}$.
Using $\ell_q'=e^{\gamma+\psi}$, we obtain $1-e^\gamma\lambda\tau/q=e^{\psi/(q-1)}$. The Legendre relation $w=\lambda(\tau\ell_q'-\ell_q)$ then yields the auxiliary potential
\begin{equation}
w_q(\psi)
=q e^\psi+(1-q)e^{q\psi/(q-1)}-1.
\end{equation}

The derivatives relevant to the causality conditions are
\begin{equation*}
\begin{aligned}
w_q'
&=q\left(e^\psi-e^{q\psi/(q-1)}\right),\\
w_q''-w_q'
&=\frac{q}{1-q}e^{q\psi/(q-1)},
\qquad
w_q''(0)=\frac{q}{1-q}.
\end{aligned}
\end{equation*}
The sign of $w_q''-w_q'$ is therefore fixed by $q$.
For $\gamma\geq0$, the causal branches are
\begin{equation}
\begin{aligned}
0<q<1 &:\qquad \lambda>0,\\
q>1   &:\qquad \lambda<0.
\end{aligned}
\end{equation}

At $q=1$, the boundary function reduces to that of ModMax, and the irrelevant flow vanishes.

  We introduce $s=e^{-\psi}$ and $b=1/(1-q)$. The auxiliary potential, its derivative, and the ratio $r=w_q/w_q'$ then take the form
\begin{equation*}
\begin{aligned}
w_q&=\frac{q}{s}+(1-q)s^{b-1}-1,\\
w_q'&=\frac{q(1-s^b)}{s},\\
r(s)&=\frac{q+(1-q)s^b-s}{q(1-s^b)}.
\end{aligned}
\end{equation*}

Consequently, the physical inverse $s_q(r)$ is determined by the root in $0<s<1$ of
\begin{equation}
\bigl(1-q+qr\bigr)s^{1/(1-q)}
-s+q(1-r)=0,
\qquad 
r \in
\begin{cases}
(0,1), & 0<q<1, \\[4pt]
\left(0,\dfrac{q-1}{q}\right), & q>1.
\end{cases}
\end{equation}
The rearranged equation also admits $s=1$, but this root is excluded for $r>0$ because the original expression for $r(s)$ is undefined at $s=1$. The vacuum value is recovered by continuity. The resulting flow function is
\begin{equation}
F_q(r)
=\frac{r\,s_q(r)}
{4q\left[1-s_q(r)^{1/(1-q)}\right](1+r^2)},
\qquad
F_q(0)=\frac{1-q}{8q}.
\end{equation}
Through the invariant variable $\rho$, this determines
the corresponding stress-tensor flow:
\begin{equation*}
\partial_\lambda\mathcal L_q
=X\,\mathcal F_q(\rho),
\qquad
\mathcal F_q(\rho)
=F_q\left(\frac{\rho}{\sqrt{4-\rho^2}}\right).
\end{equation*}

For $0<q<1$, the ratio $r(\psi)=w_q/w_q'$ is strictly increasing throughout the physical auxiliary interval. Indeed, setting $a=q/(1-q)>0$, we find
\begin{equation}
(w_q')^2-w_q w_q''
=q e^{(1-a)\psi}
\left[e^{a\psi}+a e^{-\psi}-(a+1)\right]>0
\end{equation}
for $\psi>0$. The expression in brackets vanishes at the vacuum and has a positive derivative for $\psi>0$. Consequently, $dr/d\psi>0$, and since $r$ increases from $0$ to $1$, the reconstructed flow function $F_q(r)$ is single-valued throughout $0<r<1$.

The choice $q=\tfrac12$ recovers the generalized Born--Infeld model discussed above, with
\begin{equation*}
\begin{aligned}
w_{1/2}(\psi)&=\cosh\psi-1,\\
F_{1/2}(r)&=\frac{1-r^2}{8(1+r^2)},\\
\mathcal F_{1/2}(\rho)&=\frac18-\frac{\rho^2}{16}.
\end{aligned}
\end{equation*}

As another example, consider $q=\tfrac34$, for which the auxiliary potential and its relevant derivative are
\begin{equation*}
w_{3/4}(\psi)
=\frac34 e^\psi+\frac14 e^{-3\psi}-1,
\qquad
w_{3/4}''-w_{3/4}'=3e^{-3\psi}.
\end{equation*}

For $q=\tfrac34$, removing the extraneous vacuum root reduces the inverse equation to
\begin{equation}
s^3+s^2+s=\frac{3(1-r)}{1+3r}.
\end{equation}
The left-hand side is strictly increasing for $0<s<1$, ensuring a unique physical root $s(r)$ for every $0<r<1$.

The physical root can be written explicitly as
\begin{equation}
s(r)=\frac{1}{3}
\left[K(r)-\frac{2}{K(r)}-1\right],
\end{equation}
where
\begin{equation*}
K(r)=
\sqrt[3]{
\frac{7}{2}
+\frac{81(1-r)}{2(1+3r)}
+18\sqrt{\frac{6-8r+3r^2}{(1+3r)^2}}
}.
\end{equation*}

The corresponding flow function is therefore
\begin{equation}
F_{3/4}(r)
=\frac{r\,s(r)}
{3\bigl[1-s(r)^4\bigr](1+r^2)}.
\end{equation}

For $q=\tfrac32$, the CH function is
\begin{equation}
\ell_{3/2}(\tau)
=\frac{1-\left(1-\frac23 e^\gamma\lambda\tau\right)^{3/2}}
{\lambda}.
\end{equation}
Its second derivative is positive for $\lambda<0$, corresponding to the causal branch of this model. The auxiliary potential and the relevant causality combination are
\begin{equation}
w_{3/2}(\psi)
=\frac32 e^\psi-\frac12 e^{3\psi}-1,
\qquad
w_{3/2}''-w_{3/2}'=-3e^{3\psi}<0.
\end{equation}

Introducing $u=e^\psi>1$, we obtain
\begin{equation}
r(u)=\frac{u^2+u-2}{3(u^2+u)},
\qquad 0<r<\frac13.
\end{equation}
Solving for $u$ gives the physical inverse
\begin{equation}
u(r)=\frac12
\left[
\sqrt{\frac{3(3-r)}{1-3r}}-1
\right].
\end{equation}

The corresponding flow function is
\begin{equation}
F_{3/2}(r)
=-\frac{(1-3r)^2}{144(1+r^2)}
\left[
3+\sqrt{\frac{3(3-r)}{1-3r}}
\right],
\qquad 0<r<\frac13.
\end{equation}

\subsection{No-$\tau$-maximum theory}

The no-$\tau$-maximum theory is related to the $q=\tfrac32$ model through the coupling redefinition
$\lambda_{3/2}=-\tfrac32\lambda_{\mathrm{NM}}$. Its CH boundary function is
\begin{equation*}
\ell_{\mathrm{NM}}(\tau)
=\frac{2}{3\lambda}
\left[
\left(1+e^\gamma\lambda\tau\right)^{3/2}-1
\right].
\end{equation*}
The corresponding auxiliary potential is
\begin{equation*}
w_{\mathrm{NM}}(\psi)
=\frac13e^{3\psi}-e^\psi+\frac23,
\end{equation*}
which satisfies $w_{3/2}=-\tfrac32 w_{\mathrm{NM}}$.

Introducing $u=e^\psi>1$, we have $w'_{\mathrm{NM}}=u^3-u$. Since the auxiliary potentials of the two theories differ only by a constant factor, their ratios coincide:
\begin{equation*}
r_{\mathrm{NM}}(u)=r_{3/2}(u)
=\frac{u^2+u-2}{3(u^2+u)}.
\end{equation*}
Consequently, they share the same physical inverse,
\begin{equation*}
u(r)=\frac12\left[
\sqrt{\frac{3(3-r)}{1-3r}}-1
\right],
\qquad 0<r<\frac13.
\end{equation*}

The corresponding flow functions are related by $F_{\mathrm{NM}}=-\tfrac32 F_{3/2}$. Hence,
\begin{equation}
F_{\mathrm{NM}}(r)
=\frac{(1-3r)^2}{96(1+r^2)}
\left[
3+\sqrt{\frac{3(3-r)}{1-3r}}
\right],
\qquad 0<r<\frac13.
\end{equation}
Moreover, $w_{\mathrm{NM}}''-w_{\mathrm{NM}}' =2e^{3\psi}>0$, confirming causality throughout the physical auxiliary interval.


\begin{table}[htbp]
\centering
\small
\renewcommand{\arraystretch}{1.5}
\begin{tabular}{@{}lccc@{}}
\toprule
Theory
& $w''-w'$
& Physical $r$-range
& \shortstack{Strong-field limit\\of $A_+$}
\\
\midrule
GBI
& $e^{-\psi}$
& $0<r<1$
& $0$
\\
Logarithmic
& $1$
& $0<r<1$
& $0$
\\
No-$\tau$-maximum
& $2e^{3\psi}$
& $0<r<\frac13$
& $\frac12$
\\
$q$-deformed, $0<q<1$
& $\dfrac{q}{1-q}e^{-q\psi/(1-q)}$
& $0<r<1$
& $0$
\\
$q$-deformed, $q>1$
& $\dfrac{q}{1-q}e^{-q\psi/(1-q)}$
& $0<r<\frac{q-1}{q}$
& $\dfrac{1}{2q-1}$
\\
\bottomrule
\end{tabular}
\caption{Comparison of causal models in terms of
$w''-w'$, the physical range of $r$, and the
strong-field limit of $A_+$.}
\label{tab:models}
\end{table}

Among these models, generalized Born--Infeld is
distinguished by a constant ratio $A_-/A_+$.
The logarithmic theory admits an inverse expressed
through the Lambert function, whereas the
$q$-deformed family includes the GBI model at
$q=\tfrac12$ and an explicit solution in radicals
at $q=\tfrac34$.

The $q$-deformed family has causal branches for
$0<q<1$ with $\lambda>0$ and for $q>1$ with
$\lambda<0$. In the former branch, the strictly
increasing ratio $r(\psi)$ ensures a single-valued
flow function throughout $0<r<1$. The $q=\tfrac32$
model and its no-$\tau$-maximum counterpart share
the restricted range $0<r<\tfrac13$.
Table~\ref{tab:models} compares the corresponding
causality quantities, physical ranges, and
strong-field limits of $A_+$.


\section{Discussion}
\label{sec:discussion}


In this paper, we have developed non-perturbative $T\bar T$-like flows in causal self-dual electrodynamics using the auxiliary-field formalism. The central structural result is that two commuting flows, an irrelevant flow with coupling $\lambda$ and a marginal ModMax flow with parameter $\gamma$, act on the single-variable potential $W(\varphi)=\lambda^{-1}w(\varphi-\gamma)$ through rescaling and translation. The corresponding Lagrangian flows take the autonomous form $\partial_\lambda\mathcal L=X\,F(r)$ and $\partial_\gamma\mathcal L=\tfrac14\sqrt{4X-Y}$ in terms of the stress-tensor invariants.

The flow function $F(r)$ and the auxiliary potential $w(\psi)$ provide two descriptions of the same theory, related by the parametric formulas \eqref{eq:kernel} and
\eqref{eq:wrec}--\eqref{eq:psirec}. These relations allow the flow to be reconstructed from the potential and, conversely, the potential to be obtained from a prescribed flow.


In the auxiliary formulation of Russo and Townsend~\cite{Russo:2025fuc,Russo:2024dual},
weak-field causality implies strong-field causality, and causality is equivalent to a condition on the second derivative of the potential. Our analysis further shows that, on the regular increasing branch, causality is expressed in terms of the flow function by the first-order inequality $D(r)>1/(1-r)$ and implies the stronger bound $0<r<1-e^{-\psi}$. The resulting range $0<r<1$ ensures the dominant and strong energy conditions, with the auxiliary ratio acquiring the physical interpretation $r=(\varepsilon-p_\perp)/(\varepsilon+p_\perp)$.

The condition $w'^2-ww''>0$, required for an autonomous, globally single-valued  and differentiable flow in $r$, is additional to causality. Causal theories that violate this condition cannot be described by a single-valued differentiable flow function of $r$ alone, since the stress-tensor ratio no longer provides a globally invertible coordinate on the theory. Characterizing such theories and determining whether their flows can be closed by including additional invariants remain natural open problems.


Another important result is the distinction between evenness of the potential about its vacuum, invariance under $\varphi$-parity, and analyticity in the Lorentz invariants $S$ and $P^2 $. For $\gamma=0$, these three properties are equivalent for analytic germs. For $\gamma>0$, however, the ModMax term itself is nonanalytic, and the  $\varphi$-parity dual of a causal theory acquires an acausal vacuum.

In the Maxwell-vacuum case, the first odd coefficient, $F'(0)$, controls the leading possible nonanalytic interaction, $(S^2+P^2)^{3/2}$. Odd powers of $\rho=\sqrt{Y/X}$ produce half-integer powers of $X,Y$, whereas even powers can still yield non-polynomial terms such as $Y^2/X$. Thus, analyticity of the Lagrangian in $S,P^2$ and polynomial dependence of the flow on $X,Y$ are distinct properties.


All reconstructed theories remain self-dual, and the Hamiltonian flow has the opposite sign to the corresponding Lagrangian flow. The characteristic coefficients $A_\pm$ are determined explicitly by $D(r)$ and $\psi(r)$. In particular, generalized Born--Infeld is uniquely characterized by a constant ratio $A_-/A_+$, providing a characterization of this theory directly in terms of the flow function.

Section~\ref{sec:models} provides the exact flow functions for generalized Born--Infeld, logarithmic, $q$-deformed, and no-$\tau$-maximum electrodynamics, while Table~\ref{tab:models} compares their causality properties and strong-field behavior. For $0<q<1$, the ratio $r$ is monotonic throughout the physical branch. For $q>1$, its physical range is restricted to $0<r<(q-1)/q$, and the strong-field limit of $A_+$ is $1/(2q-1)$ rather than zero. The no-$\tau$-maximum theory shares the $q=\tfrac32$ range $0<r<\tfrac13$ and the corresponding limit $A_+\to\tfrac12$.


Several extensions are suggested by these results. The auxiliary-field formalism also provides a framework for relating integrable sigma models and root-$T\bar T$ flows to four-dimensional duality-invariant electrodynamics \cite{Baglioni:2025tsc,Babaei-Aghbolagh:2025uoz,Babaei-Aghbolagh:2025hlm}. Interestingly, the reconstruction relations and single-valuedness criteria obtained here have direct counterparts in two dimensions \cite{Babaei-Aghbolagh:2026gbr}.

It would be worthwhile to investigate analogous questions in causal chiral two-form electrodynamics~\cite{Russo:2025wph,Ferko:2024zth} and higher-derivative extensions of ModMax~\cite{Kuzenko:2024hd}. The characteristic-curve approach to $T\bar T$ flows and self-dual electrodynamics \cite{Hou:2022csf,Chen:2025ndc}, together with applications of the CH construction to black-hole thermodynamics \cite{Babaei-Aghbolagh:2025tim}, provides further settings in which the single-valuedness criterion could be investigated.


\begin{acknowledgments}
We thank Song He, Jorge G. Russo, Dmitri Sorokin, and Roberto Tateo for their interest in this work and for fruitful discussions. We are particularly grateful to Dmitri Sorokin for his valuable insights, generous feedback, and encouraging discussions throughout this project. The work of H.B.-A. was carried out as part of the Postdoctoral Program \emph{Exploring TT-bar Deformations: Quantum Field Theory and Applications}, sponsored by Ningbo University.
\end{acknowledgments}

\bibliographystyle{JHEP}
\bibliography{Ref_AFF}

@article{Bialynicki-Birula:1984daz,
    author = "Bialynicki-Birula, I.",
    title = "{Nonlinear electrodynamics: Variations on a theme by Born and Infeld}",
    year = "1984"
}

@article{Gibbons:1995cv,
  author = {Gibbons, G.~W. and Rasheed, D.~A.},
  title = {Electric–magnetic duality rotations in nonlinear electrodynamics},
  journal = {Nucl. Phys. B},
  volume = {454},
  pages = {185--206},
  year = {1995},
  eprint = {hep-th/9506035}
}

@article{Gaillard:1981rj,
  author = {Gaillard, M.~K. and Zumino, B.},
  title = {Duality rotations for interacting fields},
  journal = {Nucl. Phys. B},
  volume = {193},
  pages = {221--244},
  year = {1981}
}

@article{Babaei-Aghbolagh:2022uij,
    author = "Babaei-Aghbolagh, H. and Velni, Komeil Babaei and Yekta, Davood Mahdavian and Mohammadzadeh, H.",
    title = "{Emergence of non-linear electrodynamic theories from TT{\textasciimacron}-like deformations}",
    eprint = "2202.11156",
    archivePrefix = "arXiv",
    primaryClass = "hep-th",
    reportNumber = "IPM/P-2022/13",
    doi = "10.1016/j.physletb.2022.137079",
    journal = "Phys. Lett. B",
    volume = "829",
    pages = "137079",
    year = "2022"
}

@article{Borsato:2022tmu,
    author = "Borsato, Riccardo and Ferko, Christian and Sfondrini, Alessandro",
    title = "{Classical integrability of root-TT{\textasciimacron} flows}",
    eprint = "2209.14274",
    archivePrefix = "arXiv",
    primaryClass = "hep-th",
    doi = "10.1103/PhysRevD.107.086011",
    journal = "Phys. Rev. D",
    volume = "107",
    number = "8",
    pages = "086011",
    year = "2023"
}

@article{Ferko:2022cix,
    author = "Ferko, Christian and Sfondrini, Alessandro and Smith, Liam and Tartaglino-Mazzucchelli, Gabriele",
    title = "{Root-$T \bar T$ Deformations in Two-Dimensional Quantum Field Theories}",
    eprint = "2206.10515",
    archivePrefix = "arXiv",
    primaryClass = "hep-th",
    doi = "10.1103/PhysRevLett.129.201604",
    journal = "Phys. Rev. Lett.",
    volume = "129",
    number = "20",
    pages = "201604",
    year = "2022"
}

@article{Conti:2022egv,
    author = "Conti, Riccardo and Romano, Jacopo and Tateo, Roberto",
    title = "{Metric approach to a $ \mathrm{T}\overline{\mathrm{T}} $-like deformation in arbitrary dimensions}",
    eprint = "2206.03415",
    archivePrefix = "arXiv",
    primaryClass = "hep-th",
    doi = "10.1007/JHEP09(2022)085",
    journal = "JHEP",
    volume = "09",
    pages = "085",
    year = "2022"
}

@article{Ferko:2024ali,
    author = "Ferko, Christian and Smith, Liam",
    title = "{Infinite Family of Integrable Sigma Models Using Auxiliary Fields}",
    eprint = "2405.05899",
    archivePrefix = "arXiv",
    primaryClass = "hep-th",
    doi = "10.1103/PhysRevLett.133.131602",
    journal = "Phys. Rev. Lett.",
    volume = "133",
    number = "13",
    pages = "131602",
    year = "2024"
}

@article{Bielli:2024ach,
    author = "Bielli, Daniele and Ferko, Christian and Smith, Liam and Tartaglino-Mazzucchelli, Gabriele",
    title = "{Integrable higher-spin deformations of sigma models from auxiliary fields}",
    eprint = "2407.16338",
    archivePrefix = "arXiv",
    primaryClass = "hep-th",
    doi = "10.1103/PhysRevD.111.066010",
    journal = "Phys. Rev. D",
    volume = "111",
    number = "6",
    pages = "066010",
    year = "2025"
}

@article{Bielli:2025uiv,
    author = "Bielli, Daniele and Ferko, Christian and Galli, Michele and Tartaglino-Mazzucchelli, Gabriele",
    title = "{Higher-Spin Currents and Flows in Auxiliary Field Sigma Models}",
    eprint = "2504.17294",
    archivePrefix = "arXiv",
    primaryClass = "hep-th",
    month = "4",
    year = "2025"
}

@article{Smirnov:2016lqw,
  author = {Smirnov, F.~A. and Zamolodchikov, A.~B.},
  title = {On space of integrable quantum field theories},
  journal = {Nucl. Phys. B},
  volume = {915},
  pages = {363--383},
  year = {2017},
  eprint = {1608.05499}
}

@article{Cavaglia:2016oda,
  author = {Cavaglià, Andrea and Negro, Stefano and Szécsényi, István M. and Tateo, Roberto},
  title = {$T\bar T$-deformed 2D quantum field theories},
  journal = {JHEP},
  volume = {10},
  pages = {112},
  year = {2016},
  eprint = {1608.05534}
}

@article{Russo:2024xnh,
    author = "Russo, Jorge G. and Townsend, Paul K.",
    title = "{Causality and energy conditions in nonlinear electrodynamics}",
    eprint = "2404.09994",
    archivePrefix = "arXiv",
    primaryClass = "hep-th",
    doi = "10.1007/JHEP06(2024)191",
    journal = "JHEP",
    volume = "06",
    pages = "191",
    year = "2024"
}

@article{Russo:2025wph,
    author = "Russo, Jorge G. and Townsend, Paul K.",
    title = "{Causal chiral 2-form electrodynamics}",
    eprint = "2504.01467",
    archivePrefix = "arXiv",
    primaryClass = "hep-th",
    month = "4",
    year = "2025"
}

@article{Russo:2025fuc,
    author = "Russo, Jorge G. and Townsend, Paul K.",
    title = "{Simplified self-dual electrodynamics}",
    eprint = "2505.08869",
    archivePrefix = "arXiv",
    primaryClass = "hep-th",
    month = "5",
    year = "2025"
}

@article{Bandos:2020jsw,
    author = "Bandos, Igor and Lechner, Kurt and Sorokin, Dmitri and Townsend, Paul K.",
    title = "{A non-linear duality-invariant conformal extension of Maxwell's equations}",
    eprint = "2007.09092",
    archivePrefix = "arXiv",
    primaryClass = "hep-th",
    doi = "10.1103/PhysRevD.102.121703",
    journal = "Phys. Rev. D",
    volume = "102",
    pages = "121703",
    year = "2020"
}

@article{Russo:2024csd,
  author         = "Russo, Jorge G. and Townsend, Paul K.",
  title          = "{Causal Self-Dual Electrodynamics}",
  journal        = "Phys. Rev. D",
  volume         = "109",
  pages          = "105023",
  year           = "2024",
  doi            = "10.1103/PhysRevD.109.105023",
  eprint         = "2401.06707"
}

@article{Russo:2024dual,
  author         = "Russo, Jorge G. and Townsend, Paul K.",
  title          = "{Dualities of Self-Dual Nonlinear Electrodynamics}",
  note           = "arXiv preprint",
  year           = "2024",
  eprint         = "2407.02577"
}

@article{Ferko:2023pr,
  author         = "Ferko, Christian and Kuzenko, Sergei M. and Smith, Liam and Tartaglino-Mazzucchelli, Gabriele",
  title          = "{Duality-Invariant Non-linear Electrodynamics and Stress Tensor Flows}",
  journal        = "Phys. Rev. D",
  volume         = "108",
  pages          = "106021",
  year           = "2023",
  doi            = "10.1103/PhysRevD.108.106021",
  eprint         = "2309.04253"
}

@article{Kuzenko:2024hd,
  author         = "Kuzenko, Sergei M. and Raptakis, Emmanouil S. N.",
  title          = "{Higher-derivative deformations of the ModMax theory}",
  note           = "arXiv preprint",
  year           = "2024",
  eprint         = "2404.09108"
}

@book{Courant:Hilbert,
  author         = "Courant, Richard and Hilbert, David",
  title          = "{Methods of Mathematical Physics, Vol.~II}",
  publisher      = "Wiley‐Interscience",
  address        = "New York",
  year           = "1962"
}

@article{Babaei-Aghbolagh:2024uqp,
    author = "Babaei-Aghbolagh, H. and He, Song and Ouyang, Hao",
    title = "{Generalized $ T\overline{T} $-like deformations in duality-invariant nonlinear electrodynamic theories}",
    eprint = "2407.03698",
    archivePrefix = "arXiv",
    primaryClass = "hep-th",
    doi = "10.1007/JHEP09(2024)137",
    journal = "JHEP",
    volume = "09",
    pages = "137",
    year = "2024"
}

@article{Jiang:2019epa,
    author = "Jiang, Yunfeng",
    title = "{A pedagogical review on solvable irrelevant deformations of 2D quantum field theory}",
    eprint = "1904.13376",
    archivePrefix = "arXiv",
    primaryClass = "hep-th",
    reportNumber = "CERN-TH-2019-058",
    doi = "10.1088/1572-9494/abe4c9",
    journal = "Commun. Theor. Phys.",
    volume = "73",
    number = "5",
    pages = "057201",
    year = "2021"
}

@article{He:2025ppz,
    author = "He, Song and Li, Yi and Ouyang, Hao and Sun, Yuan",
    title = "{$T\overline{T}$ Deformation: Introduction and Some Recent Advances}",
    eprint = "2503.09997",
    archivePrefix = "arXiv",
    primaryClass = "hep-th",
    month = "3",
    year = "2025"
}

@article{Ivanov:2003uj,
    author = "Ivanov, E. A. and Zupnik, B. M.",
    title = "{New approach to nonlinear electrodynamics: Dualities as symmetries of interaction}",
    eprint = "hep-th/0303192",
    archivePrefix = "arXiv",
    doi = "10.1134/1.1842299",
    journal = "Phys. Atom. Nucl.",
    volume = "67",
    pages = "2188--2199",
    year = "2004"
}

@article{Ferko:2026eci,
    author = "Ferko, Christian and Luke Martin, Cian and Sharma, Pranat",
    title = "{Auxiliary Field Deformations of the Lambda Model}",
    eprint = "2609.18831",
    archivePrefix = "arXiv",
    primaryClass = "hep-th",
    month = "9",
    year = "2026"
}

@article{Bielli:2026ggs,
    author = "Bielli, Daniele and Ferko, Christian and Galli, Michele and Tartaglino-Mazzucchelli, Gabriele",
    title = "{The classical Yangian symmetry of Auxiliary Field Sigma Models}",
    eprint = "2605.18213",
    archivePrefix = "arXiv",
    primaryClass = "hep-th",
    month = "5",
    year = "2026"
}

@article{Baglioni:2025tsc,
    author = "Baglioni, Nicola and Bielli, Daniele and Galli, Michele and Tartaglino-Mazzucchelli, Gabriele",
    title = "{Relating auxiliary field formulations of 4d duality-invariant and 2d integrable field theories}",
    eprint = "2512.21982",
    archivePrefix = "arXiv",
    primaryClass = "hep-th",
    doi = "10.1007/JHEP07(2026)039",
    journal = "JHEP",
    volume = "07",
    pages = "039",
    year = "2026"
}

@article{Fukushima:2024nxm,
    author = "Fukushima, Osamu and Yoshida, Kentaroh",
    title = "{4D Chern-Simons theory with auxiliary fields}",
    eprint = "2407.02204",
    archivePrefix = "arXiv",
    primaryClass = "hep-th",
    reportNumber = "RIKEN-iTHEMS-Report-24, STUPP-24-270",
    doi = "10.1007/JHEP09(2025)001",
    journal = "JHEP",
    volume = "09",
    pages = "001",
    year = "2025"
}

@article{Fukushima:2025tlj,
    author = "Fukushima, Osamu and Matsumoto, Takaki and Yoshida, Kentaroh",
    title = "{The Courant-Hilbert construction in 4D Chern-Simons theory}",
    eprint = "2509.22080",
    archivePrefix = "arXiv",
    primaryClass = "hep-th",
    reportNumber = "RIKEN-iTHEMS-Report-25, STUPP-25-288",
    doi = "10.1007/JHEP01(2026)122",
    journal = "JHEP",
    volume = "01",
    pages = "122",
    year = "2026"
}

@article{Fukushima:2026gan,
    author = "Fukushima, Osamu and Matsumoto, Takaki and Yoshida, Kentaroh",
    title = "{Courant-Hilbert deformations of Yang-Baxter sigma models}",
    eprint = "2602.04662",
    archivePrefix = "arXiv",
    primaryClass = "hep-th",
    reportNumber = "RIKEN-iTHEMS-Report-26, STUPP-25-293",
    doi = "10.1007/JHEP06(2026)084",
    journal = "JHEP",
    volume = "06",
    pages = "084",
    year = "2026"
}

@article{Babaei-Aghbolagh:2025uoz,
    author = "Babaei-Aghbolagh, H. and Chen, Bin and He, Song",
    title = "{Root-TT{\textasciimacron} flows unify 4D duality-invariant electrodynamics and 2D integrable sigma models}",
    eprint = "2507.22808",
    archivePrefix = "arXiv",
    primaryClass = "hep-th",
    doi = "10.1103/1r4p-3r5q",
    journal = "Phys. Rev. D",
    volume = "112",
    number = "10",
    pages = "L101702",
    year = "2025"
}

@article{Babaei-Aghbolagh:2025hlm,
    author = "Babaei-Aghbolagh, H. and Chen, Bin and He, Song",
    title = "{Integrable sigma models and universal root $T\overline{T }$ deformation via Courant-Hilbert approach}",
    eprint = "2509.17075",
    archivePrefix = "arXiv",
    primaryClass = "hep-th",
    doi = "10.1007/JHEP01(2026)108",
    journal = "JHEP",
    volume = "01",
    pages = "108",
    year = "2026"
}

@article{Babaei-Aghbolagh:2025cni,
  author = "Babaei-Aghbolagh, Hossein and Babaei Velni, Komeil and He, Song and Pezhman, Zahra",
  title = "{Root-$ T\overline{T} $ deformations on causal self-dual electrodynamic theories}",
  eprint = "2504.10361",
  archivePrefix = "arXiv",
  primaryClass = "hep-th",
  doi = "10.1007/JHEP07(2025)227",
  journal = "JHEP",
  volume = "07",
  pages = "227",
  year = "2025"
}

@article{Russo:2024kto,
    author = "Russo, Jorge G. and Townsend, Paul K.",
    title = "{Born again}",
    eprint = "2401.04167",
    archivePrefix = "arXiv",
    primaryClass = "hep-th",
    doi = "10.21468/SciPostPhys.16.5.124",
    journal = "SciPost Phys.",
    volume = "16",
    number = "5",
    pages = "124",
    year = "2024"
}

@article{Babaei-Aghbolagh:2026vkm,
  author = "Babaei-Aghbolagh, H. and Babaei Velni, Komeil and He, Song and Pezhman, Zahra",
  title = "{Classifying Causal Nonlinear Electrodynamics via $\varphi$-Parity and Irrelevant Deformations}",
  eprint = "2602.03426",
  archivePrefix = "arXiv",
  primaryClass = "hep-th",
  month = "2",
  year = "2026"
}

@article{Kuzenko:2025gvn,
    author = "Kuzenko, Sergei M. and Ruhl, Jonah",
    title = "{Generalisations of the Russo-Townsend formulation}",
    eprint = "2511.20051",
    archivePrefix = "arXiv",
    primaryClass = "hep-th",
    doi = "10.1016/j.physletb.2026.140258",
    journal = "Phys. Lett. B",
    volume = "874",
    pages = "140258",
    year = "2026"
}

@article{Kuzenko:2026hbw,
    author = "Kuzenko, Sergei M. and Ruhl, Jonah",
    title = "{Causal self-dual nonlinear electrodynamics from the Born-Infeld theory}",
    eprint = "2605.06193",
    archivePrefix = "arXiv",
    primaryClass = "hep-th",
    month = "5",
    year = "2026"
}

@article{Chen:2025ndc,
  author = "Chen, Bin and He, Song and Hou, Jue",
  title = "{Self-dual electrodynamics via the characteristic method: relativistic and Carrollian perspectives}",
  eprint = "2506.03129",
  archivePrefix = "arXiv",
  primaryClass = "hep-th",
  doi = "10.1007/JHEP02(2026)055",
  journal = "JHEP",
  volume = "02",
  pages = "055",
  year = "2026"
}

@article{Hou:2022csf,
  author = "Hou, Jue",
  title = "{$ T\overline{T} $ flow as characteristic flows}",
  eprint = "2208.05391",
  archivePrefix = "arXiv",
  primaryClass = "hep-th",
  doi = "10.1007/JHEP03(2023)243",
  journal = "JHEP",
  volume = "03",
  pages = "243",
  year = "2023"
}

@article{Babaei-Aghbolagh:2025tim,
    author = "Babaei-Aghbolagh, H. and Babaei Velni, Komeil and He, Song and Isapour, Fateme",
    title = "{Unified causal framework for a nonlinear electrodynamics black hole from the Courant-Hilbert approach: Thermodynamics and singularity}",
    eprint = "2511.17407",
    archivePrefix = "arXiv",
    primaryClass = "hep-th",
    doi = "10.1103/zfc2-52md",
    journal = "Phys. Rev. D",
    volume = "113",
    number = "8",
    pages = "084041",
    year = "2026"
}

@article{Babaei-Aghbolagh:2020kjg,
  author = "Babaei-Aghbolagh, H. and Babaei Velni, Komeil and Yekta, Davood Mahdavian and Mohammadzadeh, H.",
  title = "{$ T\overline{T} $-like flows in non-linear electrodynamic theories and S-duality}",
  eprint = "2012.13636",
  archivePrefix = "arXiv",
  primaryClass = "hep-th",
  reportNumber = "IPM/P-2020/066",
  doi = "10.1007/JHEP04(2021)187",
  journal = "JHEP",
  volume = "04",
  pages = "187",
  year = "2021"
}

@article{Ferko:2022iru,
  author = "Ferko, Christian and Smith, Liam and Tartaglino-Mazzucchelli, Gabriele",
  title = "{On Current-Squared Flows and ModMax Theories}",
  eprint = "2203.01085",
  archivePrefix = "arXiv",
  primaryClass = "hep-th",
  doi = "10.21468/SciPostPhys.13.2.012",
  journal = "SciPost Phys.",
  volume = "13",
  number = "2",
  pages = "012",
  year = "2022"
}

@article{Ferko:2023ruw, year="2023", author="Ferko, Christian and Smith, Liam and Tartaglino-Mazzucchelli, Gabriele", title="{Stress tensor flows, birefringence in non-linear electrodynamics, and supersymmetry}", eprint="2301.10411", archivePrefix="arXiv"}

@article{Babaei-Aghbolagh:2024hti,
    author = "Babaei-Aghbolagh, H. and He, Song and Morone, Tommaso and Ouyang, Hao and Tateo, Roberto",
    title = "{Geometric Formulation of Generalized Root-TT{\textasciimacron} Deformations}",
    eprint = "2405.03465",
    archivePrefix = "arXiv",
    primaryClass = "hep-th",
    doi = "10.1103/PhysRevLett.133.111602",
    journal = "Phys. Rev. Lett.",
    volume = "133",
    number = "11",
    pages = "111602",
    year = "2024"
}

@article{Born:1934gh,
  author = "Born, M. and Infeld, L.",
  title = "{Foundations of the new field theory}",
  doi = "10.1098/rspa.1934.0059",
  journal = "Proc. Roy. Soc. Lond. A",
  volume = "144",
  number = "852",
  pages = "425--451",
  year = "1934"
}

@article{Sorokin:2021tge,
    author = "Sorokin, Dmitri P.",
    title = "{Introductory Notes on Non-linear Electrodynamics and its Applications}",
    eprint = "2112.12118",
    archivePrefix = "arXiv",
    primaryClass = "hep-th",
    doi = "10.1002/prop.202200092",
    journal = "Fortsch. Phys.",
    volume = "70",
    number = "7-8",
    pages = "2200092",
    year = "2022"
}

@article{Zamolodchikov:2004ce, year="2004", author="Zamolodchikov, A. B.", title="{Expectation value of composite field T anti-T in two-dimensional quantum field theory}", eprint="hep-th/0401146", archivePrefix="arXiv"}

@article{Taylor:2018xcy, year="2018", author="Taylor, Marika", title="{$T\bar{T}$ deformations in general dimensions}", eprint="1805.10287", archivePrefix="arXiv"}

@article{Brizio:2026ynw,
    author = "Brizio, Nicol{\`o} and Kade, Moritz and Sfondrini, Alessandro and Sorokin, Dmitri P.",
    title = "{More on $ T\overline{T} $-like deformations in higher dimensions}",
    eprint = "2602.24058",
    archivePrefix = "arXiv",
    primaryClass = "hep-th",
    doi = "10.1007/JHEP06(2026)262",
    journal = "JHEP",
    volume = "06",
    pages = "262",
    year = "2026"
}

@article{Baglioni:2026mii,
    author = "Baglioni, Nicola and Bielli, Daniele and Shah, Parita and Tartaglino-Mazzucchelli, Gabriele",
    title = "{Integrable auxiliary field sigma model and the metric approach to $T\bar{T}$-like deformations}",
    eprint = "2610.13501",
    archivePrefix = "arXiv",
    primaryClass = "hep-th",
    month = "10",
    year = "2026"
}

@article{Fukushima:2026znr,
    author = "Fukushima, Osamu and Kurihara, Miyabi and Matsumoto, Takaki and Yoshida, Kentaroh",
    title = "{A gravitational realization of Courant--Hilbert deformations}",
    eprint = "2609.35452",
    archivePrefix = "arXiv",
    primaryClass = "hep-th",
    reportNumber = "RIKEN-iTHEMS-Report-26, STUPP-26-305",
    month = "9",
    year = "2026"
}

@article{Pezhman:2026dsr,
    author = "Pezhman, Zahra and Babaei-Aghbolagh, H. and Babaei Velni, Komeil",
    title = "{Exact $T\bar{T}$-Like Flows in Nonlinear Electrodynamics}",
    eprint = "2610.23281",
    archivePrefix = "arXiv",
    primaryClass = "hep-th",
    month = "10",
    year = "2026"
}

@article{Babaei-Aghbolagh:2026gbr,
    author = "Babaei-Aghbolagh, H. and He, Song and Pezhman, Zahra",
    title = "{Exact stress-tensor generators distinguish integrable continuations}",
    eprint = "2610.31297",
    archivePrefix = "arXiv",
    primaryClass = "hep-th",
    month = "10",
    year = "2026"
}

@article{Ferko:2024zth,
    author = "Ferko, Christian and Kuzenko, Sergei M. and Lechner, Kurt and Sorokin, Dmitri P. and Tartaglino-Mazzucchelli, Gabriele",
    title = "{Interacting chiral form field theories and $ T\overline{T} $-like flows in six and higher dimensions}",
    eprint = "2402.06947",
    archivePrefix = "arXiv",
    primaryClass = "hep-th",
    doi = "10.1007/JHEP05(2024)320",
    journal = "JHEP",
    volume = "05",
    pages = "320",
    year = "2024"
}

@article{Pezhman:2026sby,
    author = "Pezhman, Zahra and Babaei Velni, Komeil and Babaei-Aghbolagh, H.",
    title = "{Exact Stress Tensor Flows in Causal Electrodynamics: Courant-Hilbert Approach}",
    eprint = "2610.33887",
    archivePrefix = "arXiv",
    primaryClass = "hep-th",
    month = "10",
    year = "2026"
}

@article{Blair:2024aqz,
    author = "Blair, Chris D. A. and Lahnsteiner, Johannes and Obers, Niels A. and Yan, Ziqi",
    title = "{Matrix theory reloaded: a BPS road to holography}",
    eprint = "2410.03591",
    archivePrefix = "arXiv",
    primaryClass = "hep-th",
    reportNumber = "IFT-UAM/CSIC-24-139, NORDITA 2024-033",
    doi = "10.1007/JHEP02(2025)024",
    journal = "JHEP",
    volume = "02",
    pages = "024",
    year = "2025"
}

@article{article,
author = {Bialynicki-Birula, Iwo},
year = {1983},
month = {01},
pages = {},
title = {Nonlinear electrodynamics: Variations on a theme by Born and Infeld}
}
\end{document}